\documentclass[aps,prl,reprint,groupedaddress]{revtex4-1}
\usepackage{epsf}
\usepackage{epsfig}
\usepackage{graphicx}
\usepackage{color}
\usepackage{siunitx}
\usepackage{epstopdf}
\usepackage{amsmath}
\begin{document}
\title {\large
Direct Experimental Access to the Nonadiabatic Initial Momentum Offset upon Tunnel Ionization}
\author{S. Eckart$^{1}$}
\email{eckart@atom.uni-frankfurt.de}
\author{K. Fehre$^{1}$}
\author{N. Eicke$^{2}$}
\author{A. Hartung$^{1}$}
\author{J. Rist$^{1}$}
\author{D. Trabert$^{1}$}
\author{N. Strenger$^{1}$}
\author{A. Pier$^{1}$}
\author{L. Ph. H. Schmidt$^{1}$}
\author{T. Jahnke$^{1}$}
\author{M. S. Sch\"offler$^{1}$}
\author{M. Lein$^{2}$}
\author{M. Kunitski$^{1}$}
\author{R.~D\"{o}rner$^{1}$}
\email{doerner@atom.uni-frankfurt.de}
\affiliation{$^1$ Institut f\"ur Kernphysik, Goethe-Universit\"at, Max-von-Laue-Str. 1, 60438 Frankfurt, Germany \\
$^2$ Institut f\"ur Theoretische Physik, Leibniz Universit\"at Hannover, Appelstr. 2, 30167 Hannover, Germany}
\date{\today}
\begin{abstract}
We report on the non-adiabatic offset of the initial electron momentum distribution in the plane of polarization upon single ionization of argon by strong field tunneling and show how to experimentally control the degree of non-adiabaticity. 
Two-color counter- and co-rotating fields ($390$ and $\SI{780}{\nano\meter}$) are compared to show that the non-adiabatic offset strongly depends on the temporal evolution of the laser electric field. We introduce a simple method for the direct access to the non-adiabatic offset using two-color counter- and co-rotating fields. Further, for a single-color circularly polarized field at $\SI{780}{\nano\meter}$ we show that the radius of the experimentally observed donut-like distribution increases for increasing momentum in the light propagation direction. Our observed initial momentum offsets are well reproduced by the strong-field approximation (SFA). A mechanistic picture is introduced that links the measured non-adiabatic offset to the magnetic quantum number of virtually populated intermediate states.
\end{abstract}
\maketitle
Tunneling is one of the most intriguing quantum effects, which is well understood for transmission through a quasi-static barrier \cite{Keldysh1965}. Much less is known about the transmission through time-dependent potential barriers. One of the open questions is how a rotation of the tunnel's direction influences the momentum distribution of the particle which exits the tunnel. For a static tunnel (adiabatic tunneling) there is cylindrical symmetry around the tunnel direction and thus the initial momentum after tunneling must be isotropic in the plane perpendicular to the tunnel direction \cite{Ammosov1986,Delone1991}. For a rotating tunnel this symmetry is broken. This can lead to an offset momentum in (or against) the direction in which the tunnel exit evolves with time (non-adiabatic tunneling) \cite{Jivesh2013,Olga2011A}. 

Time-dependent potential barriers are routinely realized by exposing an atom to a strong femtosecond laser pulse. The joint electric field of the ionic core and the laser pulse gives rise to a potential barrier through which a bound electron can tunnel. For circularly polarized light the tunnel rotates in the polarization plane. Once in the continuum, the electron will be driven by the laser field. This will add a momentum given by the instantaneous negative vector potential $-\vec{A}(t)$ to the initial momentum the electron had at the tunnel exit $\vec{p_{i}}$. Including Coulomb interaction after tunneling, the post tunneling propagation can be precisely modeled using classical simulations \cite{Corkum1989,Shilovski2016,Ni2018_theo}. From this, one might hope that the question of a possible offset momentum upon exiting the tunnel can be answered experimentally by subtracting $-\vec{A}(t)$ from the measured electron momentum distribution. However, the vector potential is significantly larger than the expected $\vec{p_{i}}$ and in experiments the laser intensity (and thus $-\vec{A}(t)$) is hardly known with sufficient precision. Most previous attempts to experimentally prove the existence of offsets in the initial momentum distributions at the tunnel exit are therefore based on comparing theoretical predictions with experiments \cite{Arissian2010,Pfeiffer_2012,Xufei2014,Hofmann2016exp,Li2017_exp} (see \cite{Popruzhenko2008,Geng2014_ellipticallytheory,Geng2015OTCtheory,Li2016_theory,Ni2018_theo} for alternative theoretical approaches).

In the present Letter we solve this problem by two experimental approaches which allow us to keep the vector potential constant while changing the degree of non-adiabaticity. We observe, that this modifies the final momenta of the electrons significantly. This shows - almost free of theoretical modeling - that the initial tunnel exit momenta depend on the degree of non-adiabaticity. In the first approach we manipulate the angular velocity of the electric field vector using co- and counter-rotating two-color circularly polarized fields  \cite{becker1999schemes,Geng2015OTCtheory,fleischer2014spin,Mancuso2015,Eckart2016,Mancuso2016PRL,Han2017,Han2018Rabbitt}. The second approach is to select different electron momenta in the laser propagation direction for circularly polarized light, which is shown to be equivalent to changing the degree of non-adiabaticity. 

The two-color fields are generated using a $200$-$\mu\text{m}$ BBO to frequency double a $780$-$\text{nm}$ laser pulse (\mbox{KMLabs} Dragon, $40$-$\text{fs}$ FWHM, $\SI{8}{\kilo\hertz}$) using the same optical setup as in \cite{Eckart2016,EckartNatPhys2018}. A spherical mirror ($f=\SI{80}{\milli\meter}$) focuses the laser field (aperture of $\SI{8}{\milli\meter}$ ($\SI{5}{\milli\meter}$) for $\SI{780}{\nano\meter}$ ($\SI{390}{\nano\meter}$)) into an argon gas jet produced by supersonic gas expansion. The 3D electron momentum distributions from single ionization of argon presented in this work have been measured using cold-target recoil-ion momentum spectroscopy (COLTRIMS) \cite{ullrich2003recoil,jagutzki2002multiple}. The momentum spectrometer is the same as in \cite{EckartNatPhys2018}.

\begin{figure}
\epsfig{file=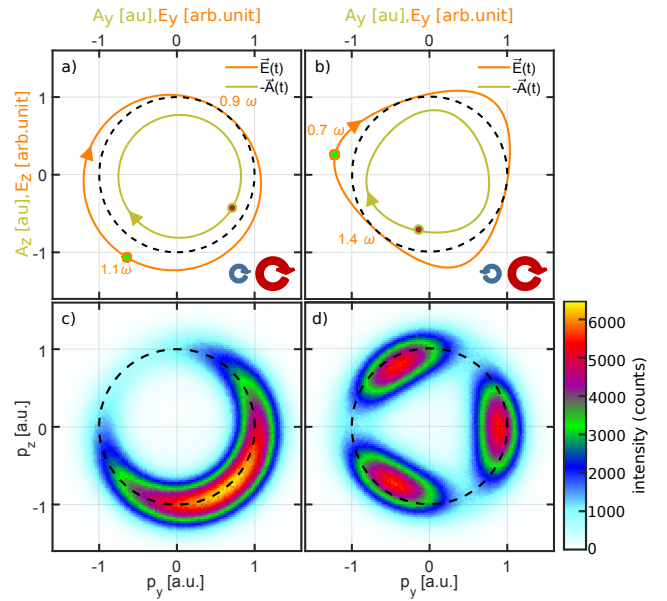, width=8.7cm} 
\caption{Two-color field composed of a strong fundamental ($E_{780}=0.046 \,$a.u.) and a weaker second harmonic ($E_{390}=0.005 \,$a.u.) field. Both wavelengths are circularly polarized and co-rotating (a) and (c) or counter-rotating (b) and (d). Depending on the helicity of the second harmonic the derivative of the electric field can be modified (phase of highest field is shown as a dot). The angular velocity of the laser electric field for the co-rotating case is $\omega_{\rm{eff}}=1.1$\,$\omega$ ($\omega_{\rm{eff}}=0.9 $\,$\omega$) and $\omega_{\rm{eff}}=0.7$\,$\omega$ ($\omega_{\rm{eff}}=1.4 $\,$\omega$) for the  counter-rotating case at the maximal (minimal) electric field. (c) and (d) show the corresponding experimental electron momentum distributions for single ionization of Ar. The black dashed circle guides the eye and has the same radius in all panels. The arrows indicate the temporal evolution of the laser electric field $\vec{E}(t)$ and the negative vector potential $-\vec{A}(t)$.}
\label{fig_figure1} 
\end{figure}

In our first experimental approach to investigate non-adiabaticity we choose the $\SI{390}{\nano\meter}$ field to be weak compared to the $\SI{780}{\nano\meter}$ field and both contributing electric fields to be circularly polarized. We switch the helicity of the second harmonic every 120 seconds to make the electric field vectors of both colors be co- or counter-rotating. Fig. \ref{fig_figure1}(a) and \ref{fig_figure1}(b) show the resulting combined laser electric fields and vector potentials. The key feature is that in both cases the minimum and maximum of the combined vector potential are the same. The maximum (minimum) is reached when the vector potentials of the two corresponding colors are parallel (antiparallel). However, a decisive parameter for the non-adiabaticity - the rotational speed of the tunnel exit - is different. The effective angular frequency at the minimum of the vector potential is $\omega_{\rm{eff,co}}=\frac{1-2\eta}{1-\eta}\omega=0.9$\,$\omega$ ($\omega_{\rm{eff,counter}}=\frac{1-2\eta}{1+\eta}\omega=0.7$\,$\omega$) for the co- (counter-) rotating field ($\omega_{\rm{eff}}=\frac{|\dot{E}|}{|E|}$ where $E$ is the instantaneous combined electric field). Here $\omega$ corresponds to $\SI{780}{\nano\meter}$ and $\eta=\frac{E_{390}}{E_{780}}$ defines the two-color field ratio (all definitions are valid for $0 \leq \eta<0.5$). Alternatively the effective, instantaneous Keldysh-parameters ($\gamma_{\rm{eff}}=\frac{\omega_{\rm{eff}}}{E}\sqrt{2  I_p}$ where $I_p$ is the ionization potential) for the time of minimal vector potential can be compared. They are $\gamma_{\rm{eff,co}}=\frac{1-2\eta}{(1-\eta)^2}\frac{\omega}{E_{780}}\sqrt{2 I_p}=1.3$ for the co-rotating and $\gamma_{\rm{eff,counter}}=\frac{1-2\eta}{(1+\eta)^2}\frac{\omega}{E_{780}}\sqrt{2 I_p}=0.9$ for the counter-rotating scenario.

Since tunneling is a highly nonlinear process, the electron most likely escapes the atomic potential at the peak electric field (see dots in Fig. \ref{fig_figure1}(a) and  \ref{fig_figure1}(b)). Fig. \ref{fig_figure1}(c) and \ref{fig_figure1}(d) show the associated electron momentum distributions (which are integrated over $|p_x|=0.0\pm0.5$\,a.u.). In the absence of non-adiabatic offsets the most probable radial momentum $p_r^{peak}(\varphi)$ for each angle $\varphi$ would be given by the absolute value of the corresponding negative vector potential. Deviations to this value are expected to be due to non-adiabatic offsets in the initial momentum distribution or Coulomb interaction of the electron with its parent ion. 

\begin{figure}
\epsfig{file=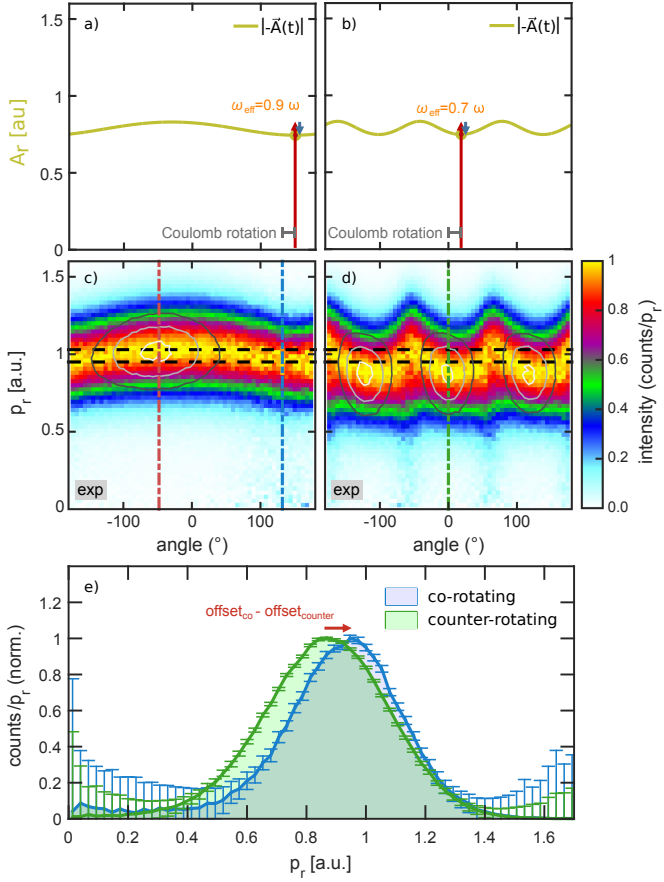, width=8.7cm} 
\caption{The golden line in (a) and (b) shows the negative vector potentials from Fig. \ref{fig_figure1} in cylindrical coordinates. The minimal vector potential is identical in (a) and (b) and marked with a long red and a short blue arrow that indicate the two driving fields that have anti-parallel vector potentials at this time. The effective angular frequency is not equal at this instant ($\omega_{\rm{eff}}=0.9$\,$\omega$ in (a) and $\omega_{\rm{eff}}=0.7$\,$\omega$ in (b)). (c)-(e) show the data from Fig. \ref{fig_figure1} in cylindrical coordinates where $p_{x}$ is the momentum component parallel to the light propagation and $p_{r}=\sqrt{p_{y}^2+p_{z}^2}$ (subset of data with  $|p_{x}|=0.15 \pm 0.05 \,$a.u.). The data in (c) and (d) has been normalized columnwise. The most probable radial momentum $p_r^{peak}(\varphi)$ is angle-dependent. The minimal and the maximal $p_r^{peak}$ in (c) are marked as black dashed lines and guide the eye in (c) and (d). (e) shows subsets of (c) and (d) restricting the angle to $132\pm\SI{10}{\degree}$ in (c) and $0\pm\SI{10}{\degree}$ in (d). The angular gates are indicated by blue and green vertical lines, which are shifted with respect to the corresponding vector potential due to the Coulomb rotation (obtained from the CTS model). Contour lines indicate the intensity in (c) and (d) prior to normalization of each angle individually. The dark gray, light gray and white lines indicate 45\%, 70\% and 95\% of maximal intensity.}
\label{fig_figure2} 
\end{figure}

The quantitative change of the initial momentum at the tunnel exit can be seen best by inspecting the data in cylindrical coordinates. Fig. \ref{fig_figure2}(a) and (b) show the negative vector potentials from Fig. \ref{fig_figure1}(a) and \ref{fig_figure1}(b) in cylindrical coordinates where $A_r=\sqrt{A_y^2+A_z^2}$.
Fig. \ref{fig_figure2}(c) and \ref{fig_figure2}(d) show the same data as in Fig. \ref{fig_figure1}(c) and \ref{fig_figure1}(d). The color scale encodes the intensity normalized to the maximum value for every angle in the plane of polarization independently. The actual intensity distribution is shown as contour lines. The plotted intensity represents the counts divided by the radial electron momentum $p_r=\sqrt{p_y^2+p_z^2}$. This takes the volume element $p_r$ into account. We note that the Coulomb field does lead to an angular shift of the distribution maximum in the polarization plane, which is indicated in Fig. \ref{fig_figure2}(a) and (b).

The minimal and the maximal value for $p_r^{peak}(\varphi)$ in the co-rotating scenario are used as references and are marked as dashed black lines to guide the eye. Inspecting Fig. \ref{fig_figure2}(d) it is obvious that the minimal radial momentum in the counter-rotating case is much lower than in Fig. \ref{fig_figure2}(c). To underline this result, Fig. \ref{fig_figure2}(e) compares the radial momentum distributions restricting the angle in the polarization plane (see vertical blue and green dashed lines). The radial electron momentum distributions in Fig. \ref{fig_figure2}(e) are shifted by about $0.1 \,$a.u. relative to each other although the corresponding negative vector potentials are the same in both cases. This allows to conclude that the offset of the initial momentum distribution in the co-rotating case is bigger by about $0.1 \,$a.u. than in the counter-rotating case.

\begin{figure}
\epsfig{file=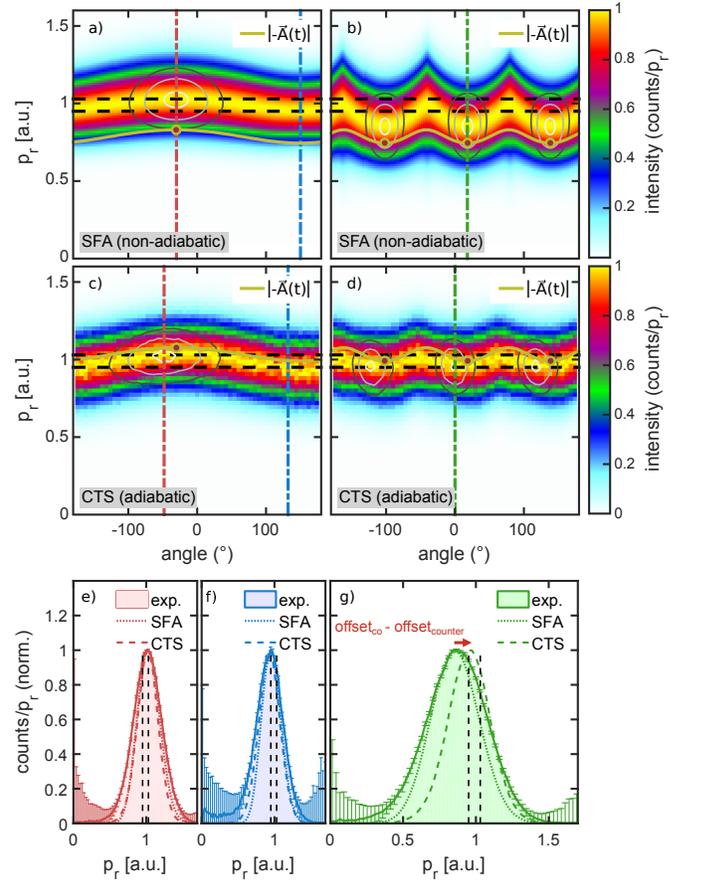, width=8.7cm} 
\caption{Theoretical modeling of the experimental data shown in Fig. \ref{fig_figure2}. The left column shows co-rotating fields, the right column counter-rotating fields. The black dashed lines from Fig. \ref{fig_figure3}(c) serve as references in all panels. (a) and (b) show the (non-adiabatic) SFA result neglecting intracycle interference. (c) and (d) show the result from the (adiabatic) CTS model. The intensity has been chosen for SFA ($E_{780}=0.046 \,$a.u. and $E_{390}=0.005 \,$a.u.) and CTS ($E_{780}=0.060 \,$a.u. and $E_{390}=0.005 \,$a.u.) independently such that (a) and (c) match Fig. \ref{fig_figure2}(c). (e) and (f) show subsets of (a) ((c) and Fig. \ref{fig_figure2}(c)) restricting the angle to $-30\pm\SI{10}{\degree}$ and $150\pm\SI{10}{\degree}$ ($-48\pm\SI{10}{\degree}$ and $132\pm\SI{10}{\degree}$). (g)  shows subsets of (b) ((d) and \ref{fig_figure2}(d)) restricting the angle to $18\pm\SI{10}{\degree}$ ($0\pm\SI{10}{\degree}$). The corresponding angular gates are indicated by colored vertical lines. The golden lines in (a)-(d) show the negative vector potentials.  Contour lines in (a)-(d) indicate the intensity  prior to normalization as in Fig. \ref{fig_figure2}. All data shown is restricted to $|p_{x}|=0.15 \pm 0.05 \,$a.u.}
\label{fig_figure3} 
\end{figure}

We emphasize that the gained insight regarding the initial non-adiabatic momenta is obtained purely from the experimental data, without the need for an exact knowledge of the intensity. In the next step we do some modeling in particular to rule out the possible concern that the observed radial momentum change could originate from differences in the Coulomb interaction for co- and counter-rotating fields. To this end we look at two complementary theoretical models in Fig. \ref{fig_figure3}. First we analyze the results from strong-field approximation (SFA), which incorporates initial momentum offsets but neglects Coulomb interaction \cite{Misha2005}. Neglecting pre-exponential factors, the momentum dependent ionization amplitude in saddle-point SFA is proportional to $\exp{\left(-iS\right)}$, with the action
\begin{align}
S & =\int_{t_s} \left[ \frac{1}{2}\left( \vec{p} +\vec{A}(t) \right)^2 +I_p  \right] dt
\end{align}
evaluated at its saddle-points $t_s$. The second is a classical two-step (CTS) model that includes Coulomb interaction but does not include any offsets in the initial momentum distribution. Here we follow the procedure described in \cite{Shilovski2016} using the potential $V(r)=-1/r$, neglecting the semiclassical phase and using a 12 cycle laser field (total duration, sine-square envelope). The CTS model incorporates the prediction of adiabatic tunneling, i.e. an initial Gaussian momentum distribution which is centered at zero in both directions perpendicular to the tunnel exit, and zero initial momentum in tunnel direction. In particular, the tunnel exit for each trajectory was obtained by solving Eq. (5) from Ref. \cite{Shilovski2012} and the ionization probability is calculated according to Eq. (9) from Ref. \cite{Shilovski2016}.

For both models we adjust field intensities of the two colors such that the predictions of the respective model regarding the minimal and maximal $p_r^{peak} (\varphi)$ match the experiment for the co-rotating scenario and we then use these intensities also for the counter-rotating case. The SFA in Fig. \ref{fig_figure3}(a)-(b) nicely reproduces the experimentally observed shift of the radial momenta upon changing the helicity of the second harmonic, while the classical model does not show such a shift. This clearly rules out Coulomb interaction as the origin of the shift and indicates that the non-adiabatic initial momenta are reliably included in SFA.

Motivated by this success of SFA we inspect the action in SFA for a driving electric field in the $p_yp_z$-plane: 
\begin{align}
S & =\int_{t_s} \left[ \frac{1}{2}   \begin{pmatrix}
  0\\
  p_y+ A_y(t)\\
  p_z+A_z(t)\\
 \end{pmatrix} ^2  +I_p +\frac{1}{2}p_x^2  \right] dt
\end{align}
We note that in SFA an increase in $p_x$ is equivalent to introducing an effective ionization potential $I_p^{\rm{eff}}=I_p+\frac{1}{2}p_x^2$. The effective Keldysh-Parameter, which is a measure for the non-adiabaticity, can be rewritten as $\gamma_{\rm{eff}}= \frac{\omega_{eff}}{E_0} \sqrt{2 I_p^{\rm{eff}}}$.

This is the idea of our second approach to change the non-adiabaticity of the tunneling-process while keeping the vector potential constant: Due to the equivalence of $I_p$ and $\frac{1}{2}p_x^2$ in SFA, the electron momentum component in the light propagation direction $p_x$ is expected to influence the non-adiabatic offset momenta. 

\begin{figure}
\epsfig{file=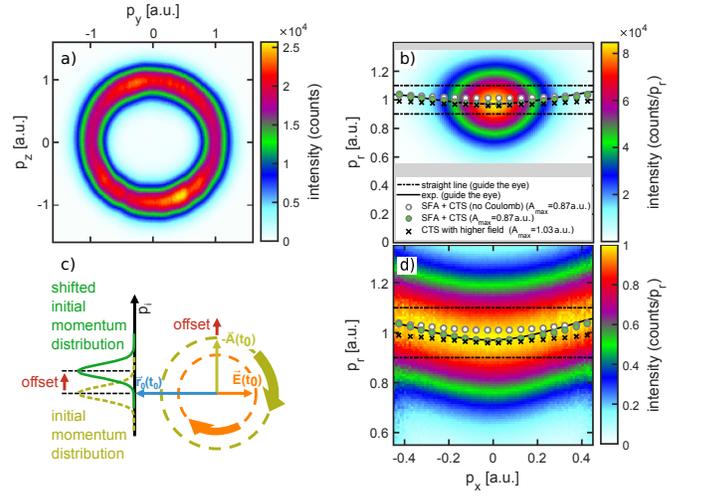, width=8.7cm} 
\caption{Influence of the electron momentum $p_x$ (along the laser propagation direction) on $p_r$ (radial electron momentum in the plane of polarization). (a) experimental electron  momentum distributions from single ionization of Ar by circularly polarized light at $\SI{780}{\nano\meter}$. (b) shows the same data as (a) but in cylindrical coordinates. (c) illustrates the geometry at the instant of tunneling $t_0$: the direction of the laser electric field $E(t_0)$, the direction of the tunnel exit $r_0(t_0)$, the non-adiabatic offset (red arrow), the negative vector potential $-A(t_0)$ and the initial momentum ($p_i$) distribution at the tunnel exit. (d) shows the same as (b) after every column is normalized individually. The maximum of each column in (b) and (d) is indicated by the black solid line to guide the eye. The horizontal golden line shows the negative vector potential. The classical two-step (CTS) simulation with a peak vector potential of $A_{max}=0.87$\,a.u. including the initial momentum offset from SFA has been calculated without (white dots) and including Coulomb interaction (green dots) of the electron with the ion after tunneling. The black crosses show the CTS simulation for increased peak vector potential ($A_{max}=1.03$\,a.u.) and without initial momentum offset but with Coulomb interaction.}
\label{fig_figure4} 
\end{figure}

To show this experimentally, we analyse the simplest possible scenario: ionization by single-color circularly polarized light. Fig. \ref{fig_figure4}(a) shows the resulting, well known donut-shaped electron momentum distribution (integrated over $|p_x|=0.0\pm0.5$\,a.u.). Fig. \ref{fig_figure4}(b) shows the same data as (a) but in cylindrical coordinates (integrating over the angle in the polarization plane). 

Fig. \ref{fig_figure4}(d) shows the data from Fig. \ref{fig_figure4}(b), with each column being normalized independently. The resulting banana-like electron momentum distribution shows that the momentum component $p_r$ strongly depends on $p_x$. If the final electron momentum were accurately described by the negative vector potential, the radius of the donut would be independent of the momentum component $p_x$. However, the experiment shows that this is not the case. 

Excellent agreement with the experiment (see green dots in Fig. \ref{fig_figure4}(b) and \ref{fig_figure4}(d)) is reached by performing the same numerical CTS simulation as above, in which we now offset the initial momentum by the $p_x$-dependent value determined from SFA momentum distributions (momentum offset $p_i(p_x)= 0.18164 + 0.12825p_x^2 - 0.0091726p_x^4$). For comparison the same calculation is done neglecting the Coulomb potential after tunneling. This shows the bare offset momentum from SFA (white dots in Fig. \ref{fig_figure4}(b) and \ref{fig_figure4}(d)). Both calculations use the same peak electric field which is chosen to fit the experiment. For comparison the black crosses in Fig. \ref{fig_figure4} show the results from the CTS model without any non-adiabatic offset and with increased intensity. It is evident, that only the full model (green data points) reaches excellent agreement with the experiment. This result shows that the dependence of the radial electron momentum $p_r$ on the momentum component in the light propagation direction $p_x$ is partly due to the initial momentum distribution introduced by the offset of the initial momentum distribution predicted by SFA and can be fully understood including Coulomb interaction. 

What is the origin of the non-adiabatic offset? In static tunneling, the potential is time-independent and energy must be conserved. In non-adiabatic tunneling this is no longer true. In SFA it can be seen that due to the time-dependent laser potential the electron gains energy during the under-the-barrier motion \cite{KlaiberPRL2015}. In another approach to this question, Klaiber and Briggs \cite{Klaiber2016} have suggested that non-adiabatic tunneling occurs in two steps. First by excitation to a virtual intermediate state by few photon absorption and second by (adiabatic) tunneling from that virtual state. For circularly polarized light the magnetic quantum number $m$ of the intermediate off-shell-state is equal to the number of virtually absorbed photons. States of positive $m$ possess a ring current co-rotating with the vector potential. Upon tunneling this leads to an increased final momentum \cite{EckartNatPhys2018,Olga2011A,Olga2011B,Liu2017_exp}. Assuming conservation of angular momentum during tunneling one can estimate the increase in $p_i$ to be given by $r_0 \times p_i= m$ \cite{EckartNatPhys2018}. In Fig. \ref{fig_figure4} we observe $p_i =0.18$\,a.u. which leads to $m=1.8$ (for $r_0 = 10$\,a.u.). This suggests that the magnetic quantum number of the virtual intermediate state is not only experimentally accessible but also the mechanistic origin of the non-adiabatic offset.
 
In conclusion, we have experimentally shown that tunneling through a rotating barrier exhibits non-adiabatic features that depend on the effective angular frequency of the laser electric field $\omega_{\rm{eff}}$ and on the effective ionization potential $I_p^{\rm{eff}}$. Higher momenta in the light propagation direction $p_x$ result in higher radial momenta in the plane of polarization $p_r$ for single-color fields. In addition to the conceptual interest of modification of the tunneling process this has practical consequences as $p_r$ is routinely used for calibration of the laser intensity \cite{Alnaser2004,Smeenk2011,Eckle2008NatPhys}. Furthermore our experimental two-color scheme comparing co- and counter-rotating fields with otherwise identical field parameters opens up new avenues to study atomic and molecular systems investigating non-adiabaticity and the momentum distribution of the initial state \cite{EckartNatPhys2018} free of the inevitable uncertainties of the laser intensity. 

\section{Acknowledgments}
\normalsize
We thank Yunquan Liu and Armin Scrinzi for helpful discussions. This work was supported by the DFG Priority Programme ``Quantum Dynamics in Tailored Intense Fields'' of the German Research Foundation (project DO 604/29-1). K.F. and A.H. acknowledge support by the German National Academic Foundation.

\bibliographystyle{apsrev4-1}

\end{document}